\documentclass[apj]{emulateapj}

\newcommand{\commentx}[1]{}

\newcommand{\ra}[3]   
   {\makebox[1.5em][r]{#1}\makebox[1.5em][r]{#2} \makebox[2em][r]{#3}}

\newcommand{\hms}[3]  
   {${#1}^{\mathrm{h}}{#2}^{\mathrm{m}}{#3}^{\mathrm{s}}$}

\newcommand{\hmin}[2]  
   {\ensuremath{{#1}^{\mathrm{h}}{#2}^{\mathrm{m}}}}
   
\newcommand{\hours}[1]  
   {\ensuremath{{#1}^{\mathrm{h}}}}

\newcommand{\dms}[3]  
   {\ensuremath{{#1}\degree{#2}\arcminute{#3}\arcsecond}}

\newcommand{\dm}[2]  
   {\ensuremath{{#1}\degree{#2}\arcminute}}

\newcommand{\ukcmb}  
           {\ensuremath{\micro \kelvin_\mathrm{cmb}}}

\newcommand{\uk}  
           {\ensuremath{\micro \kelvin}}

\newcommand{\fdeg} 
           {\hbox{$.\!\!^{\circ}$}}

\usepackage{color}

\newcommand{\beq}{\begin{equation}}
\newcommand{\eeq}{\end{equation}}
\newcommand{\be}{\begin{equation}}
\newcommand{\ee}{\end{equation}}
\newcommand{\bea}{\begin{eqnarray}}
\newcommand{\eea}{\end{eqnarray}}
\newcommand{\bdi}{\begin{displaymath}}
\newcommand{\edi}{\end{displaymath}}

\def\lsim{\,\lower2truept\hbox{${<\atop\hbox{\raise4truept\hbox{$\sim$}}}$}\,}
\def\gsim{\,\lower2truept\hbox{${>\atop\hbox{\raise4truept\hbox{$\sim$}}}$}\,}

\usepackage{natbib}  
\usepackage{subfigure}  

\shorttitle{ A search for Low Variance Circles in the CMB Sky}

\shortauthors{Amir Hajian}

\begin{document}

\title{Are there echoes from the pre-Big Bang Universe? A search for Low Variance Circles in the CMB Sky}

\author{
Amir~Hajian\altaffilmark{\dag,1}
}
\altaffiltext{1}{Canadian Institute for Theoretical Astrophysics, University of
Toronto, Toronto, ON\ M5S~3H8, Canada}
\altaffiltext{\dag}{\url{ahajian@cita.utoronto.ca}}

\begin{abstract}
The existence of concentric low variance circles in CMB sky, generated by black-hole encounters in an aeon preceding our big bang, is a prediction of the Conformal Cyclic Cosmology. Detection of three families of such circles in WMAP data was recently reported by Gurzadyan \& Penrose (2010). We reassess the statistical significance of the low variance circles detected by Gurzadyan \& Penrose by comparing with Monte Carlo simulations of the CMB sky with realistic modeling of the anisotropic noise in WMAP data.  We  find that all three groups are consistent at 3$\sigma$ with a Gaussian CMB sky as predicted by inflationary cosmology model.  
\end{abstract}

\keywords{cosmology: cosmic microwave background, cosmology: cosmology: observations -- (cosmology:) large-scale structure of universe}

\section{Introduction}
In a recent paper \cite{Gurzadyan:2010da} reported a high significance detection of concentric circles in the Cosmic Microwave Background (CMB) maps with anomalously low variance. The existence of these circles, if true, pose a serious challenge to our understanding of the CMB as being a Gaussian random field withing the framework of inflationary cosmology. The above authors  used data from  the  the Wilkinson Microwave Anisotropy Probe (WMAP) to look for concentric low variance circles in the CMB sky. They examined 10885 choices of center in the CMB sky after masking the galactic plain by excluding the $|b|<20^\circ$ region from the maps. For each choice of center, they computed the variance of the temperature fluctuations in successively larger concentric rings of $0.5^\circ$, at increasing radii. They found three groups of rings of low variance at various radii. In this paper we compare the variance of the above low-variance-circles with the average variance of Monte Carlo simulations of the CMB sky to assess the statistical significance of these circles. 

\section{Data}
We use co-added inverse-noise weighted data from seven single year maps observed by WMAP at $94$ GHz (W-band) and $61$ GHz (V-band).
The maps are foreground cleaned (using the foreground template model discussed in \cite{hinshaw/etal:2007}) and are at HEALPix \footnote{\url{http://healpix.jpl.nasa.gov}}  resolution 9 ($N_{\rm side}=512$). The WMAP data are signal dominated on large scales, $l < 548$ \citep{larson/etal:prep} and the detector noise dominates at smaller scales. The noise in WMAP data is a non-uniform (anisotropic) white noise that varies from pixel to pixel in the map. Pixel noise in each map is determined by $N_{obs}$ with the expression
\be \label{eq:noise}
\sigma = \sigma_0/\sqrt{N_{obs}},
\ee
where $N_{obs}$ is the number of observations used to construct each pixel. Regions with larger number of observations have lower noise variances. $N_{obs}$ is included in the maps supplied by LAMBDA website\footnote{\url{http://lambda.gsfc.nasa.gov}}. Scan strategy of the WMAP  is such that it spends more time scanning the ecliptic poles and hence the Northern Ecliptic Pole (NEP) and the Southern Ecliptic Pole (SEP) are the two regions with smallest noise variances in WMAP data. Two of the three groups of concetric low-variance circles of \cite{Gurzadyan:2010da} are located close to these low-noise regions.

In all of our analysis we use pixel masks to exclude foreground-contaminated regions of the sky from the analysis. We use temperature analysis KQ85 mask which masks 22\% of the sky including the galactic plane and the bright point sources.
 
\begin{figure}
\centering
\vspace{0.1cm}
\includegraphics[width=1\linewidth]{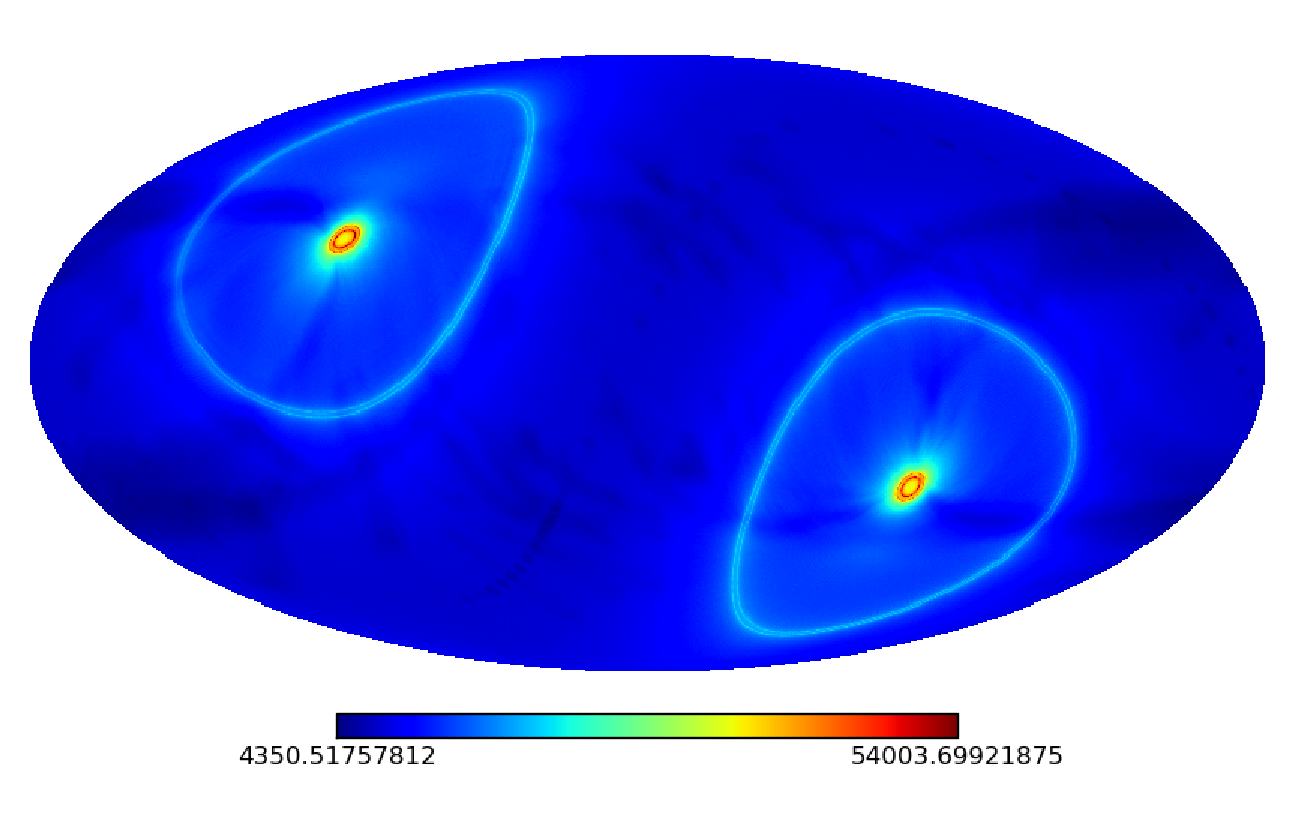}
\caption{Number of observations at every pixel, $N_{obs}$, for W-band data. Regions with larger $N_{obs}$ have smaller noise variances. This is an important feature of the WMAP data that should be taken into account in simulations of the CMB sky to correctly assess the statistical significance of the observed variances in the data.}
\label{fig:nobs}
\end{figure}
 
\section{Simulations}
In order to assess statistical significance of the results, we use Monte Carlo simulations of the CMB sky. Simulated maps have two components:
\be\label{eq:sim}
\Delta T(\hat{n}) = \Delta T_{CMB}(\hat{n})\otimes B(\hat{n}) + N(\hat{n}),
\ee
where $\Delta T_{CMB}$ is a realization of the Gaussian CMB field, $N(\hat{n})$ is the pixel noise and $\otimes B(\hat{n})$ means convolved with the proper beam of the experiment. 

We make 200 realization of the CMB sky using {\tt synfast} routine of HEALPix\footnote{We use {\tt healpy} which is a Python version of HEALPix.} with the underlying power spectrum computed with CAMB\footnote{\url{{\tt http://camb.info}}} using best fit parameters of \cite{dunkley/etal:prep}. The maps are then convolved with WMAP beams for W and V bands. Since the two frequency bands have different beam transfer functions, we make separate simulated maps for the two bands. Noise realizations are added to the beam convolved maps in the end. Noise maps are simulated using eqn.(\ref{eq:noise}) with $\sigma_0 = 6.549$ mK and $\sigma_0 =3.137$ mK for W and V-bands respectively.  We test our simulations by comparing their average power spectra with the data power spectrum.

\section{Statistic}
The statistic we use in this analysis is the variance of the temperature fluctuations along circles with a given radius, $\theta$, in the sky:
\be\label{eq:var}
V_{\hat{n}}(\theta) = \frac{1}{S_\theta}\oint_{S_\theta} (\Delta T(\hat{n'}) - \mu_{\hat{n}}(\theta))^2  \delta(\hat{n'}\cdot \hat{n} -\cos{\theta}) \rm{d}\hat{n'},
\ee
where $S_{\theta}$ is the circumference of the circle and the integral is done along the circle with a given radius, $\theta$, whose center is defined by a unit vector $\hat{n}$ on the sphere. In practice we use a discrete version of the above statistic by replacing the integral with a sum over pixels along each circle. 

\section{Results}
\begin{figure*} \label{fig}
\centering
\subfigure[W-band, $\hat{n}=(37.00^\circ,105.04^\circ)$]{\label{fig:circle1_W}\vspace{0.1cm}\includegraphics[width=.45\linewidth]{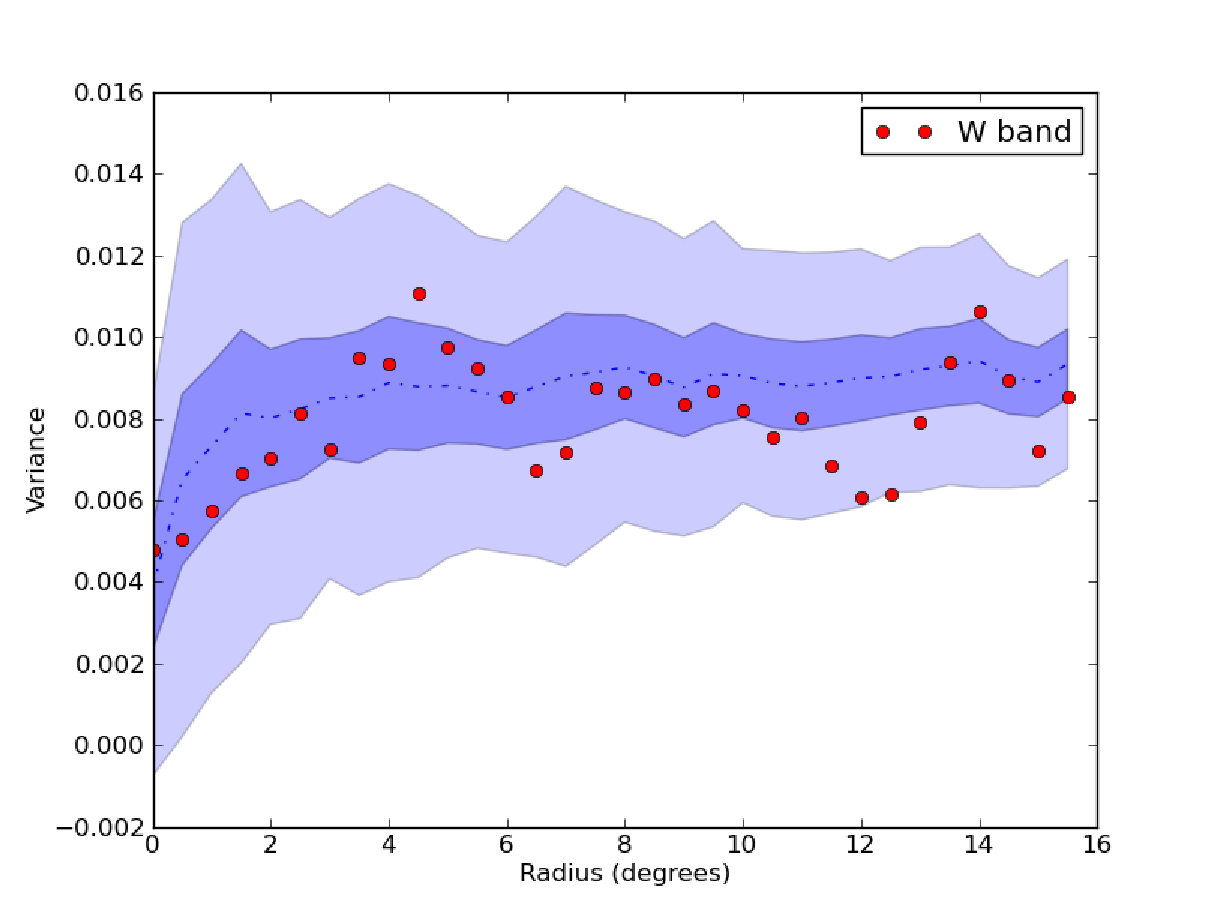}}
\subfigure[V-band, $\hat{n}=(37.00^\circ,105.04^\circ)$]{\label{fig:circle1_V}\vspace{0.1cm}\includegraphics[width=.45\linewidth]{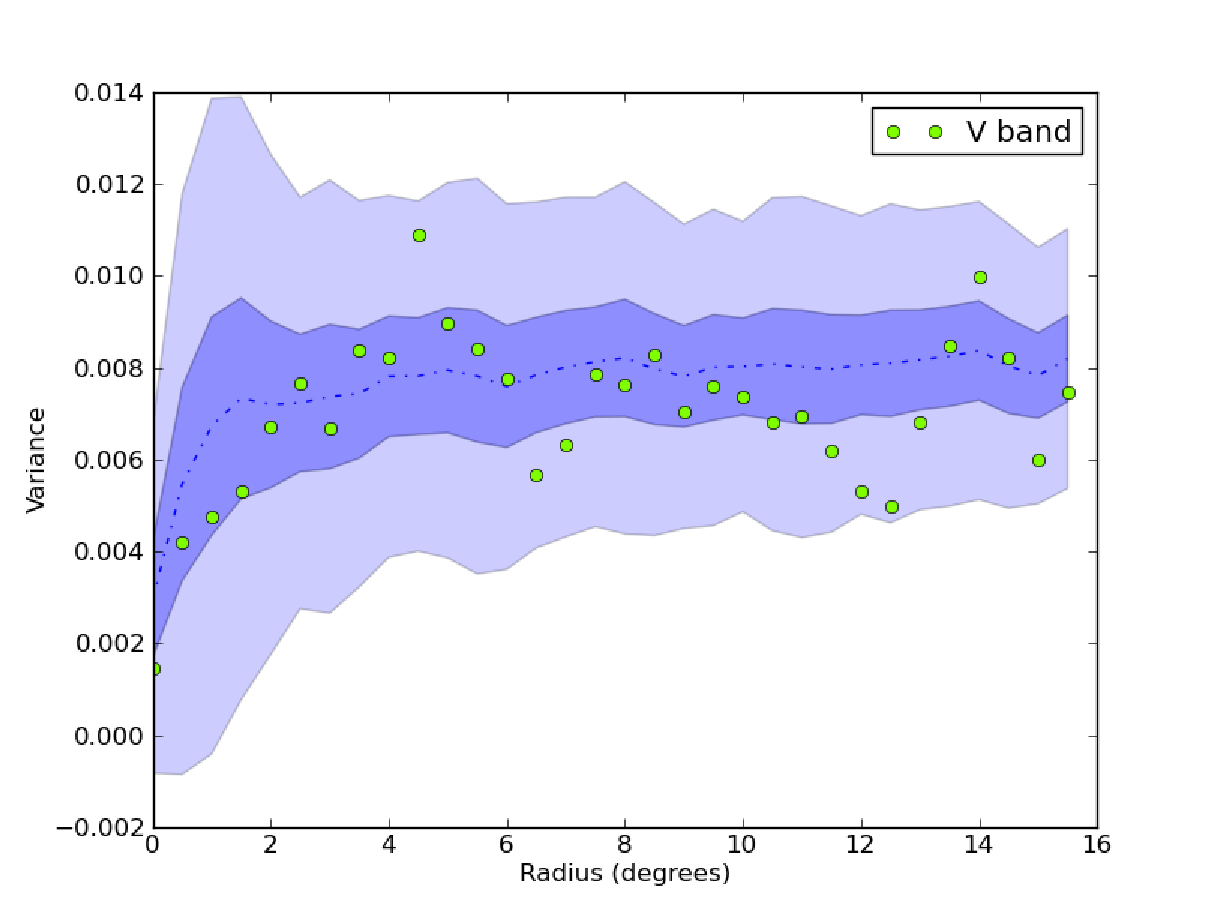}}
\subfigure[W-band, $\hat{n}=(-31.00^\circ,252.00^\circ)$]{\label{fig:circle1_V}\vspace{0.1cm}\includegraphics[width=.45\linewidth]{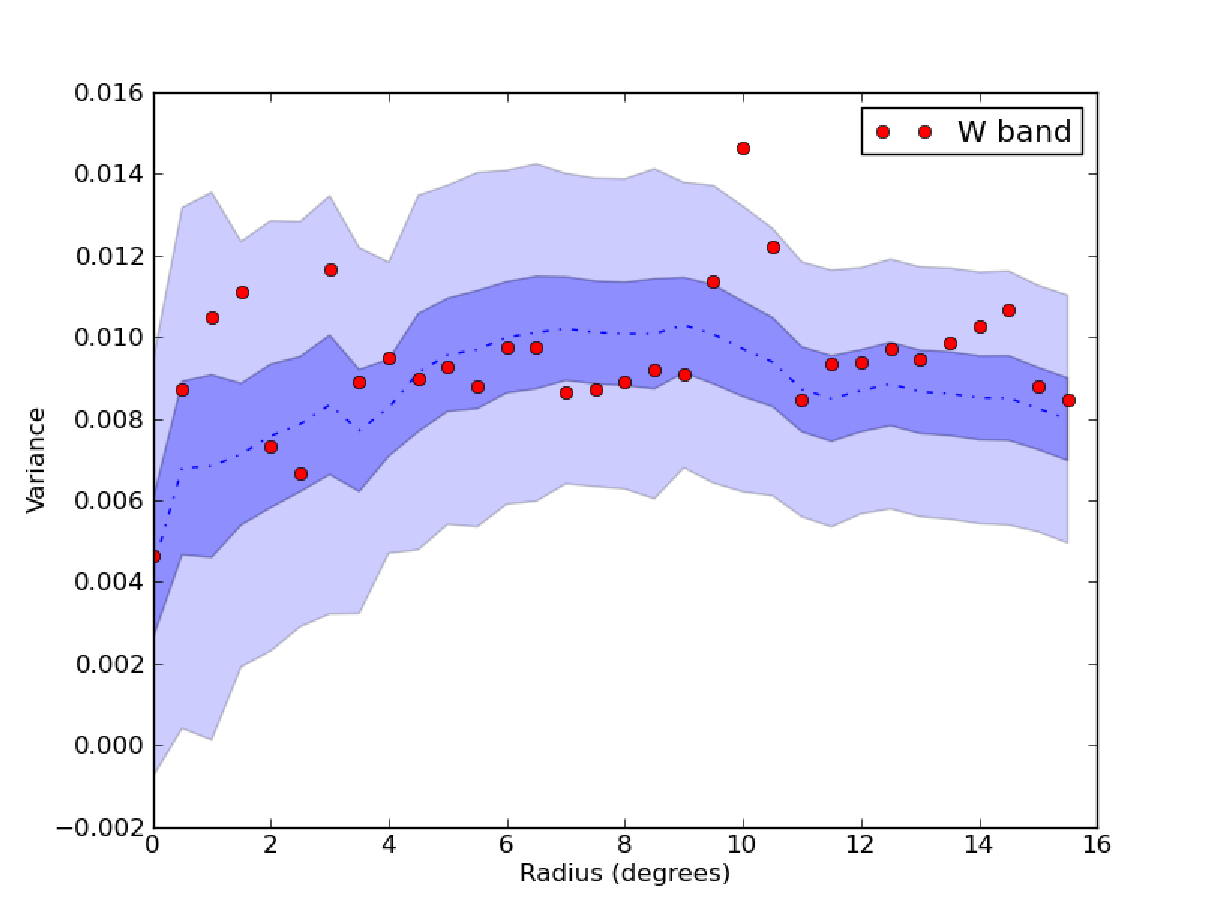}}
\subfigure[V-band, $\hat{n}=(-31.00^\circ,252.00^\circ)$]{\label{fig:circle1_V}\vspace{0.1cm}\includegraphics[width=.45\linewidth]{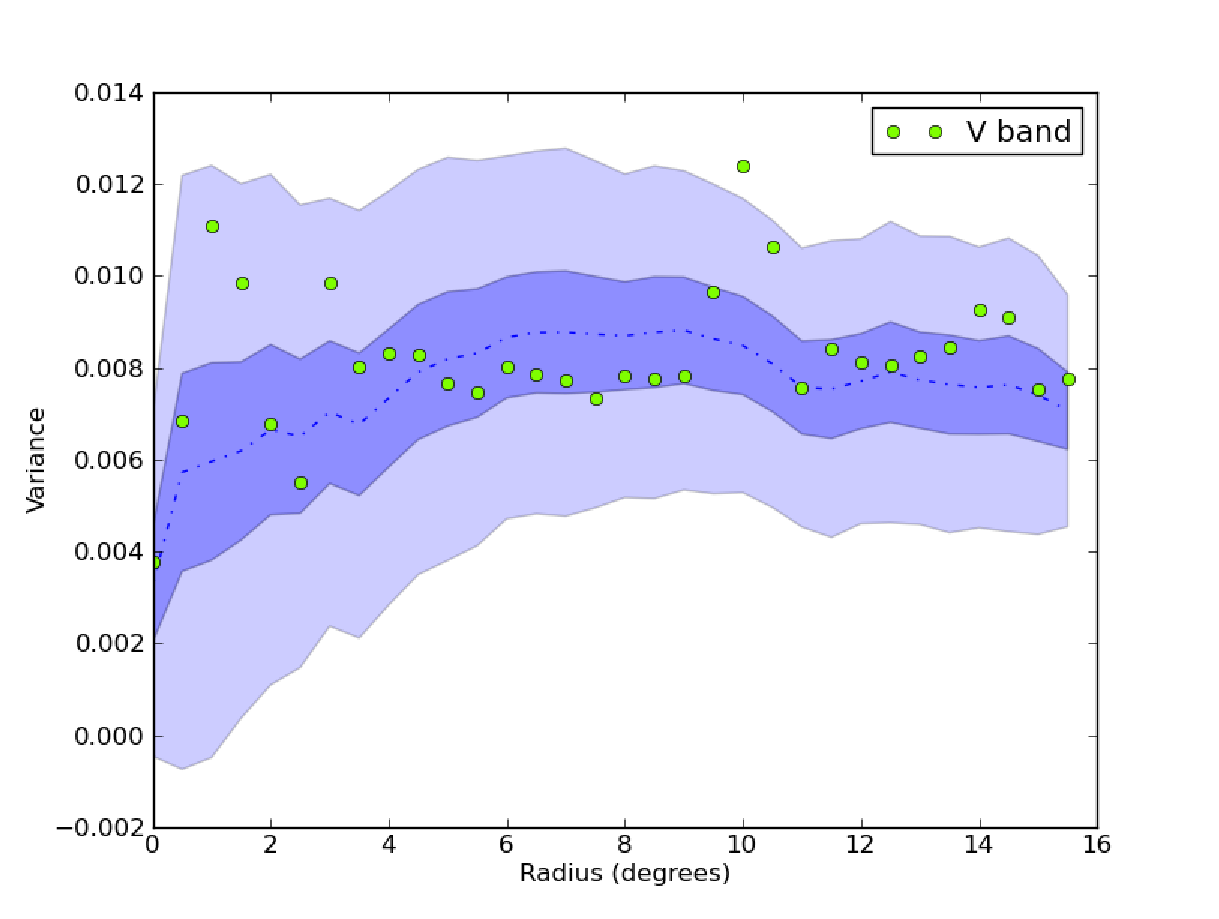}}
\subfigure[W-band, $\hat{n}=(80.25^\circ,270.00^\circ)$]{\label{fig:circle1_V}\vspace{0.1cm}\includegraphics[width=.45\linewidth]{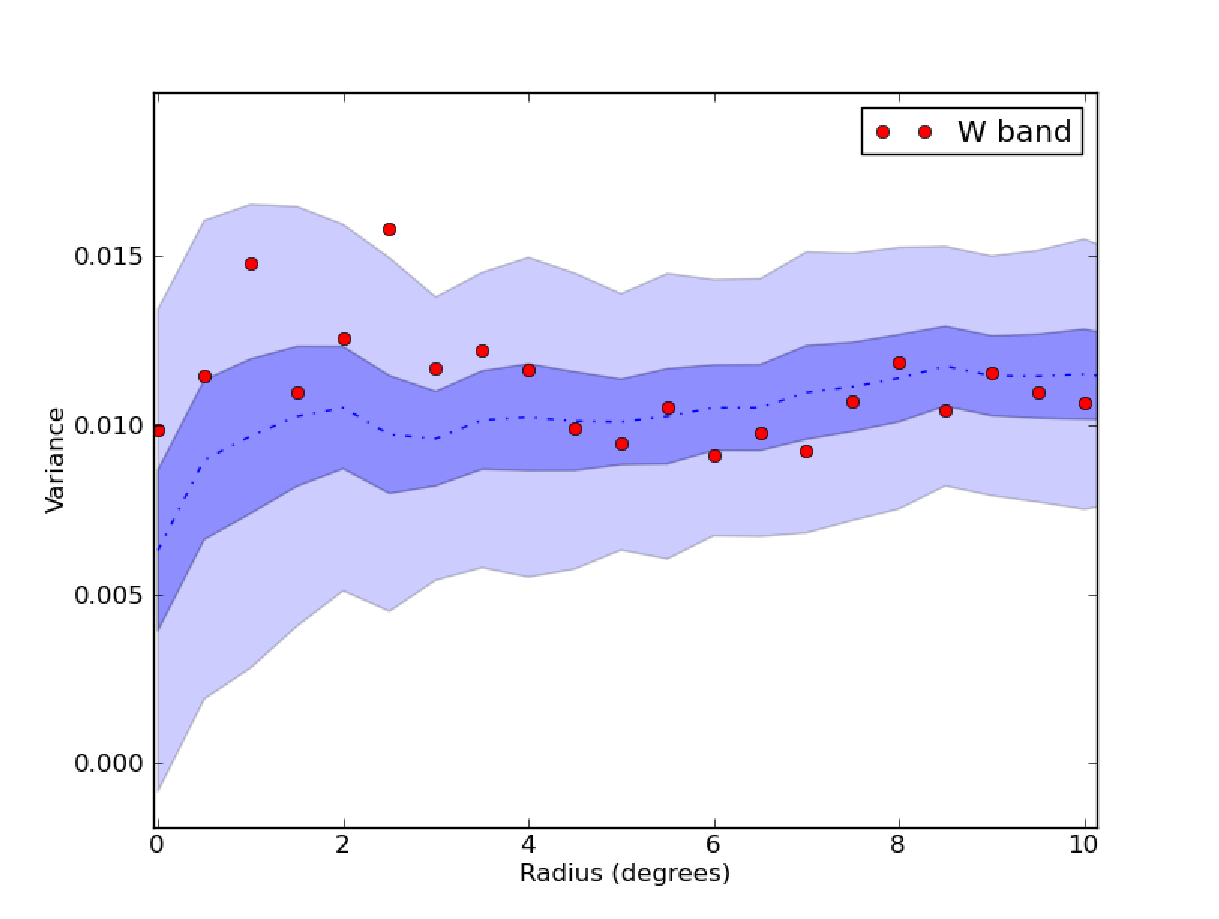}}
\subfigure[V-band, $\hat{n}=(80.25^\circ,270.00^\circ)$]{\label{fig:circle1_V}\vspace{0.1cm}\includegraphics[width=.45\linewidth]{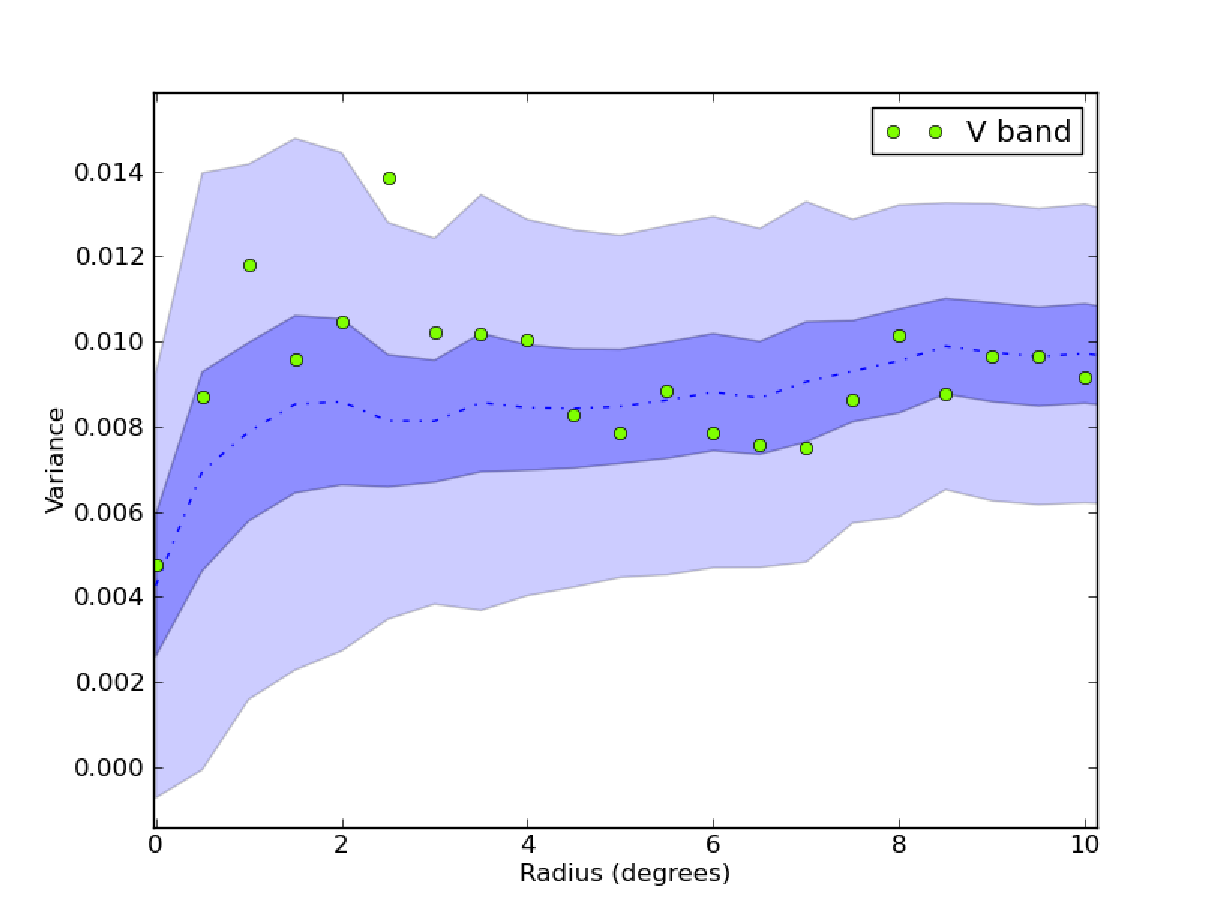}}
\caption{Variances of the low variance circles of Penrose \& Gurzadyan compared with the average variances of the same circles in 200 Monte Carlo simulations of the CMB sky with the anisotropic noise of WMAP. Left/right panels show the results of W/V bands. The dashed blue line is the mean value of the simulations of a Gaussian CMB sky with WMAP noise. Dark and light blue bands represent 1 and 3$\sigma$ levels computed from the standard deviation of the simulations. Although we see the same patterns as reported by Gurzadyan \& Penrose, none of the variances are anomalously low. These data do not support the existence of pre-big bang circles in the CMB sky.    }
\end{figure*}

\begin{figure}
\centering
\vspace{0.1cm}
\includegraphics[width=1\linewidth]{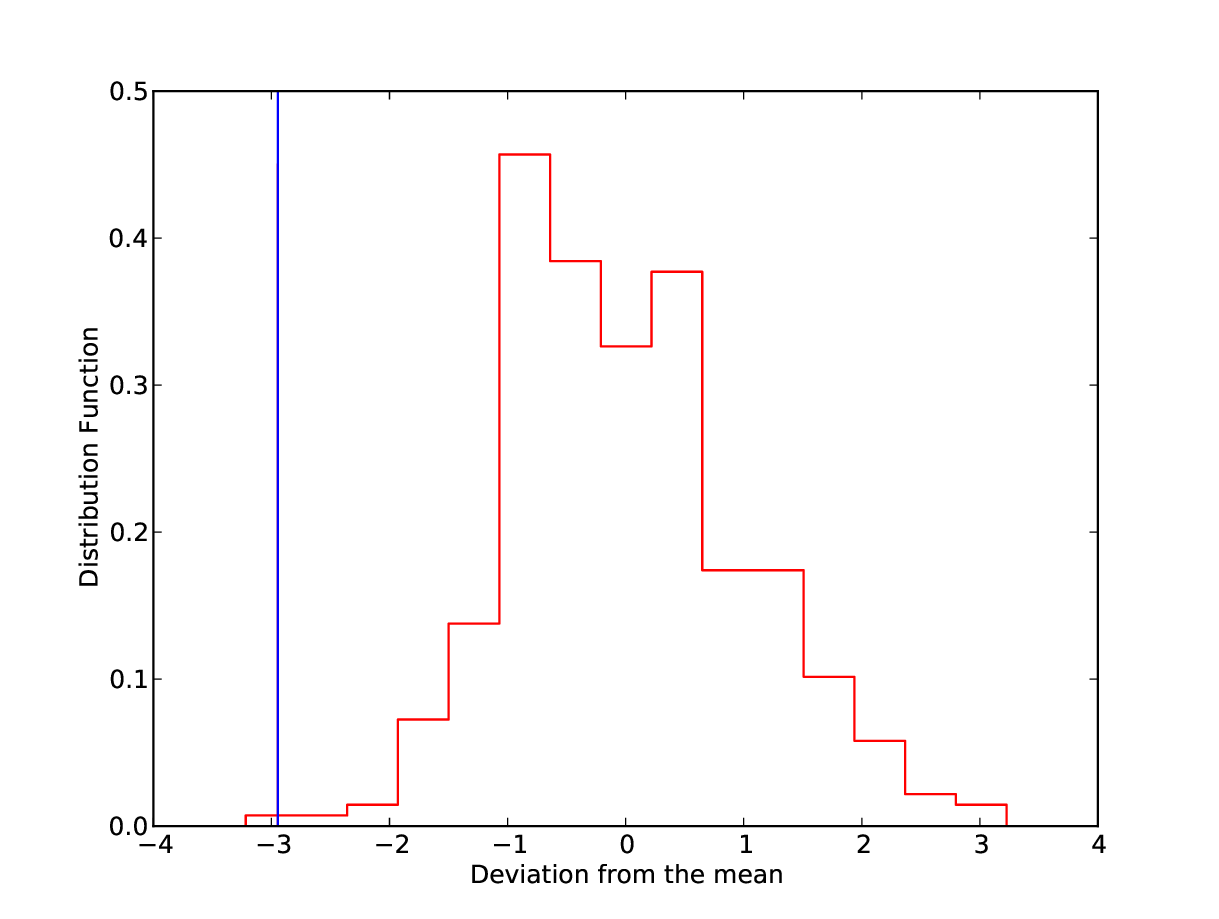}
\caption{Comparison of the most significant low variance circle, $\theta=12^\circ$, in W-band (blue line) with the normalized probability distribution function of variances of the same circle in the Monte Carlo simulations. We are plotting the standardized values on the x-axis. Even the lowest variance circle is not really anomalous when compared with the random realizations of the CMB sky that have WMAP-like anisotropic noise in them.  }
\label{fig:pdf}
\end{figure}

We apply the statistic defined in eqn.(\ref{eq:var}) on WMAP data at V- and W-bands. We choose the three families of low variance circles detected by \cite{Gurzadyan:2010da}. Centers of the three groups of rings are at $\hat{n}=(37.00^\circ,105.04^\circ)$, $(-31.00^\circ,252.00^\circ)$ and $(80.25^\circ,270.00^\circ)$. For each of these groups, we compute the variance of the temperature fluctuations of rings of $0.5^\circ$ thickness as a function of the radius of the ring. Fig. \ref{fig} shows the results. Left panels show W-band while right panels show the results from V-band. The shape of the ring variance curves are very similar to those of \cite{Gurzadyan:2010da} and we see the same peaks and troughs at the same locations and radii as reported by the above authors. We confirm the existence of the low variance circles of \cite{Gurzadyan:2010da}, but at a much lower significance. We compare the measured variances with the average of variances done in the same way on 200 simulations of the Gaussian CMB sky with WMAP noise realizations. The dashed blue line shows the average variance from the simulations at the same radii. The dark blue band is  1$\sigma$ standard deviation from the mean of the simulations and the light blue band shows the 3$\sigma$ region. As Fig. \ref{fig} shows, there is no evidence of anomalously low variance circles in WMAP data and all low-variance circles of Gurzadyan and Penrose fall below 3$\sigma$ deviation form the average of the simulated Gaussian random CMB sky. A quick estimate can give us an a lower limit on the expected number of 3$\sigma$ deviations in a map and can show how unlikely it is to get results like these in a random realization of the sky. There are 165,000 circle centers in the maps at $0.5^\circ$ smoothing and 32 rings of radius $<16^\circ$. That makes 5,000,000 possible circles at this resolution. In a Gaussian random field, we expect $5\%$ of these circles to fall beyond 2$\sigma$ (that is 250,000 occurrences in one map) and about $0.3\%$ ({\it i.e.} 15,000 circles) to be more than 3$\sigma$ away from the mean. Therefore the low-variance circles are very well consistent with random deviations from mean of the Gaussian random CMB realizations. As it is seen in Fig.  \ref{fig}, most of low variances are close to 1$\sigma$ deviation from the mean. Largest deviation happens at $\theta=12^\circ$ in W-band. In order to investigate that data point further and to make sure the non-Gaussian distribution of uncertainties does not bias our conclusion, we plot the probability distribution function (pdf) of the simulated variances and compare it with the data.  Fig. \ref{fig:pdf} shows the result. Even this extreme case is not really anomalous when compared with random realizations of the CMB sky that contain anisotropic noise similar to that of WMAP data.

\section{Summary and Conclusion} 
By comparing with Monte Carlo simulations of the CMB sky, we find that the low variance circles of  \cite{Gurzadyan:2010da} are not anomalous. They can naturally occur in a Gaussian CMB sky consistent with the predictions of the inflationary cosmology. 

\begin{acknowledgments}
I would like to thank David Spergel for his suggestions and comments throughout this work. I would also like to thank  Jim Peebles and Roger Penrose for enlightening discussions on the subject. I acknowledge the use of the Legacy Archive for Microwave Background Data Analysis (LAMBDA). Support for LAMBDA is provided by the NASA Office of Space Science.   Some of the results in this paper have been derived using the HEALPix \citep{gorski/etal:2005} package.

\end{acknowledgments}

\begin{thebibliography}{5}
\expandafter\ifx\csname natexlab\endcsname\relax\def\natexlab#1{#1}\fi

\bibitem[{{Dunkley} {et~al.}(2010){Dunkley}, {Hlozek}, {Sievers}, {Acquaviva},
  {Ade}, {Aguirre}, {Amiri}, {Appel}, {Barrientos}, {Battistelli}, {Bond},
  {Brown}, {Burger}, {Chervenak}, {Das}, {Devlin}, {Dicker}, {Bertrand
  Doriese}, {Dunner}, {Essinger-Hileman}, {Fisher}, {Fowler}, {Hajian},
  {Halpern}, {Hasselfield}, {Hernandez-Monteagudo}, {Hilton}, {Hilton},
  {Hincks}, {Huffenberger}, {Hughes}, {Hughes}, {Infante}, {Irwin}, {Juin},
  {Kaul}, {Klein}, {Kosowsky}, {Lau}, {Limon}, {Lin}, {Lupton}, {Marriage},
  {Marsden}, {Mauskopf}, {Menanteau}, {Moodley}, {Moseley}, {Netterfield},
  {Niemack}, {Nolta}, {Page}, {Parker}, {Partridge}, {Reid}, {Sehgal},
  {Sherwin}, {Spergel}, {Staggs}, {Swetz}, {Switzer}, {Thornton}, {Trac},
  {Tucker}, {Warne}, {Wollack}, \& {Zhao}}]{dunkley/etal:prep}
{Dunkley}, J., {et~al.} 2010, arXiv:1009.0866

\bibitem[{{G{\'o}rski} {et~al.}(2005){G{\'o}rski}, {Hivon}, {Banday},
  {Wandelt}, {Hansen}, {Reinecke}, \& {Bartelmann}}]{gorski/etal:2005}
{G{\'o}rski}, K.~M., {Hivon}, E., {Banday}, A.~J., {Wandelt}, B.~D., {Hansen},
  F.~K., {Reinecke}, M., \& {Bartelmann}, M. 2005, \apj, 622, 759

\bibitem[{Gurzadyan \& Penrose(2010)}]{Gurzadyan:2010da}
Gurzadyan, V.~G., \& Penrose, R. 2010, arXiv:1011.3706

\bibitem[{{Hinshaw} {et~al.}(2007){Hinshaw}, {Nolta}, {Bennett}, {Bean},
  {Dor{\'e}}, {Greason}, {Halpern}, {Hill}, {Jarosik}, {Kogut}, {Komatsu},
  {Limon}, {Odegard}, {Meyer}, {Page}, {Peiris}, {Spergel}, {Tucker}, {Verde},
  {Weiland}, {Wollack}, \& {Wright}}]{hinshaw/etal:2007}
{Hinshaw}, G., {et~al.} 2007, \apjs, 170, 288

\bibitem[{{Larson} {et~al.}(2010){Larson}, {Dunkley}, {Hinshaw}, {Komatsu},
  {Nolta}, {Bennett}, {Gold}, {Halpern}, {Hill}, {Jarosik}, {Kogut}, {Limon},
  {Meyer}, {Odegard}, {Page}, {Smith}, {Spergel}, {Tucker}, {Weiland},
  {Wollack}, \& {Wright}}]{larson/etal:prep}
{Larson}, D., {et~al.} 2010, arXiv:1001.4635

\end{thebibliography}

\end{document}